\documentclass[aps,pre,reprint,showkeys,superscriptaddress]{revtex4-2}
\usepackage{amsfonts,amssymb,amsmath,dcolumn,bm,float} 
\usepackage{graphicx,color,epsfig}
\usepackage[utf8]{inputenc}
\usepackage[T1]{fontenc}
\usepackage{mathptmx}

\begin{document}
\title{Chaotic Griffiths phase in neuron map networks}
\author{Juan Velez-Rojas}
\affiliation{Okinawa Institute of Science and Technology Graduate University, Okinawa, Japan}
\author{M. G. Cosenza}
\affiliation{School of Physical Sciences and Nanotechnology, Grupo Interdisciplinario de Sistemas Complejos, Universidad Yachay Tech, Urcuqu\'i, Ecuador}
\date{October, 2026}

\begin{abstract}
Griffiths phases provide a mechanism for sustaining critical dynamics over extended parameter regions without fine-tuning to an isolated transition point. This behavior has been mainly associated with structural heterogeneity in complex networks.
Here we show that quenched heterogeneity in local parameters can generate 
a chaotic Griffiths phase in a network of map-based neurons.  We study globally coupled Chialvo maps
with quenched disorder in a local neuronal parameter,  thereby isolating intrinsic parameter disorder from topological disorder. As the coupling strength is varied, a chaotic Griffiths phase  emerges between desynchronized and coherent chaotic dynamics, characterized by intermittent formation and breakup of neuronal clusters and large fluctuations in the fraction of clustered neurons. Within this regime, cluster sizes exhibit power-law distributions with  coupling-dependent exponents, providing evidence of critical behavior over a finite parameter interval. The chaotic Griffiths phase is absent in the homogeneous limit and broadens progressively as the degree of parameter heterogeneity increases. We further introduce topological disorder through small-world connectivity and find that its combined action with local parameter heterogeneity substantially extends the chaotic Griffiths phase. In both globally coupled and small-world networks, the number of positive Lyapunov exponents displays anomalous scaling with system size. These results demonstrate that intrinsic dynamical heterogeneity constitutes an independent mechanism for generating chaotic Griffiths phases and that its interplay with structural disorder enhances extended criticality in neuronal networks.
\end{abstract}
\maketitle

\section{Introduction} 
Networks of coupled dynamical elements exhibit a wide variety of collective phenomena, including spatiotemporal chaos, synchronization, clustering, multistability, and chimera states \cite{Boccaletti2006,Arenas00893,Parastesh0211}. 
Such collective states arise
in many physical, biological, and social systems, where both the structure of the interactions and the diversity of the constituent units strongly determine global behavior.

Collective behavior is especially relevant in neuronal systems because 
it supports essential brain functions, including information processing, memory, and cognition \cite{varela2001brainweb, buzsaki2004neuronal,fell2011role}.
Experimental observations reveal that neuronal activity exhibits complex fluctuations across a wide range of spatial and temporal scales, often characterized by scale-invariant statistics, long-range temporal correlations, and intermittent transitions between coherent and incoherent states \cite{linkenkaer2001long, Beggs2003, Petermann2009, Kitzbichler2009, tagliazucchi2012criticality}.
These observations have motivated the hypothesis that neural systems operate near criticality, where information transmission, dynamic range, and sensitivity to external stimuli 
may be enhanced \cite{Kinouchi2006, shew2009neuronal, shew2011information, Cocchi2017,Calvo2024}.

The conventional picture of criticality requires fine-tuning to a critical point separating ordered and disordered phases. Such a condition may be difficult to maintain in biological systems given their intrinsic variability.
An alternative scenario is provided by the Griffiths phase, originally introduced in the context of disordered magnetic systems \cite{Griffiths1969}. In systems with quenched disorder, rare regions may remain locally ordered even when the bulk is disordered, leading to anomalous relaxation and critical-like behavior over an extended region of parameter space rather than being restricted to a critical point \cite{Bray1987, Vojta2006,sanchez2023heterogeneity}.
The Griffiths phase is found  across multiple systems \cite{Vazquez2009, liu2021observation, amaral2021criticality}, and  is closely associated with models of complex networks \cite{Munoz2010,Vazquez2011,Cota2016,Cota2018}.
In recent years, the Griffiths phase has attracted considerable attention in 
the study of neuronal dynamics and brain networks, where heterogeneous, modular, and hierarchical network architectures have been shown to stretch criticality over broad parameter intervals \cite{Moretti2013}, thereby
avoiding the need for fine tuning
and providing robustness to changes in a control parameter.
Structurally induced Griffiths phases in neuronal models have been proposed for understanding
some critical regimes observed experimentally \cite{girardi2016griffiths, odor2016critical},
potentially accounting for 
the reported scale-invariant activity patterns in the brain \cite{Beggs2003, Kitzbichler2009, tagliazucchi2012criticality, fusca2023brain}. 
 Such states have been associated with enhanced dynamic
range and information-processing capabilities
 \cite{ Kinouchi2006,shew2011information, Beggs2008}, and are correlated with health \cite{zimmern2020brain,Shi2022,fusca2023brain}, and  cortical novelty \cite{ribeiro2026critical}.

A related phenomenon, called the chaotic Griffiths phase, was discovered
by Shinoda and Kaneko
in random networks of coupled chaotic maps \cite{Shinoda2016}. The chaotic Griffiths phase is characterized by intermittent formation and destruction of synchronized clusters, power-law distributions of cluster sizes, and anomalous scaling of Lyapunov spectra. In this regime, chaotic dynamics coexists with critical-like behavior over an extended parameter range.

Most previous investigations of the Griffiths phase in neuron systems have focused on topological disorder or structural heterogeneity, while dynamical heterogeneity has received comparatively little attention. Real neuronal systems can exhibit chaotic behavior \cite{korn2003there, hirata2012chaos} and are intrinsically heterogeneous: neurons differ in excitability, firing thresholds, and other dynamical properties \cite{marder2006variability, padmanabhan2010intrinsic, cembrowski2019heterogeneity}. 
Such diversity can be represented as quenched disorder in the local neuronal parameters.
A relevant open question is whether local heterogeneity alone, in the absence of topological disorder, can generate 
the characteristic signatures of
the chaotic Griffiths phase.

In this article, we investigate this question in a network of coupled map-based neurons with heterogeneous local parameters. 
As local dynamics, we employ the Chialvo map, which reproduces essential neuronal behaviors, including excitability, spiking, bursting, and chaos \cite{chialvo1995generic}.
In Section 2, we present a globally coupled map network model with local parameter heterogeneity,  thereby isolating intrinsic parameter
disorder from topological disorder. We show the emergence of a chaotic Griffiths phase, along with coherent and desynchronized behaviors as the coupling strength is varied.
In Section 3, we introduce statistical quantities to characterize the chaotic Griffiths phase and probe its criticality. Our results show that intrinsic parameter disorder alone is sufficient to generate the chaotic Griffiths phase.
In Section 4, 
we include  small-world
connectivity and show that topological disorder significantly extends the parameter range in which the chaotic Griffiths phase is observed.
In both cases, we find the anomalous scaling of the number of positive Lyapunov exponents  associated to the chaotic Griffiths phase, although with scaling exponents smaller than those reported for topologically disordered networks of identical maps.

These results indicate
that intrinsic and structural heterogeneity can play
distinct but complementary roles in sustaining extended
criticality in neuronal systems.
In Section 5, we present the Conclusions.

\section{Chaotic Griffiths phase in networks with heterogeneous neurons}
As local neuron dynamics, we consider the Chialvo two-dimensional map that displays excitable dynamics  \cite{chialvo1995generic}:
\begin{eqnarray}
\label{eq:chialvox}
 x_{t+1} &=&  f(x_t,y_t) = x_t^{2} \exp(y_t - x_t) + k, \\
y_{t+1}  &=& g(x_t,y_t) = a y_t - b x_t + c,
\label{eq:chialvoy}
\end{eqnarray}
where the variable $x_t$ corresponds to the activation potential that represents the voltage state of the neuron, while the variable $y_t$  expresses the recovery behavior
(like ion channel inactivation) that allows the neuron to return to its resting state after a spike. The parameters $a$, $b$, and $c$ govern the time scale and sensitivity of the recovery process, and $k$ describes the external stimulus.

We consider a globally coupled neuron map network defined as
\begin{eqnarray}
\label{eq:first_equation}
x_{t+1}(i) &=& (1-\epsilon) f\left(x_t(i), y_t(i), k(i) \right) + \epsilon h_t , \\
y_{t+1}(i) &=& g(x_t(i),y_t(i)) , \\
h_t &=& \frac{1}{N}\sum_{j=1}^{N} f\left(x_t(j), y_t(j),  k(j) \right),
\label{eq:second_equation}
\end{eqnarray}
where  $x_t(i)$,   $y_t(i)$, give the state of the $i$-th map ($i =1, \ldots, N$) at a discrete time $t$, the functions $f(x_t,y_t)$ and $g(x_t,y_t)$ are defined in Eqs.~(\ref{eq:chialvox})-(\ref{eq:chialvoy}), and the parameter $\epsilon$ represents the strength of the global coupling to the mean field $h_t$ of the
local activation variables $x_t(i)$.
We introduce heterogeneity in the neurons through the parameter $k$, by implementing a uniform distribution $k(i) \in [0.026, 0.03]$. Then, each neuron $i$ is assigned a parameter value $k(i)$ in this range, while the other local parameters $a$, $b$, and $c$ are the same for all neurons. We fix the parameter values at $a=0.89$, $b=0.18$, and $c=0.28$, 
for which the map Eqs.~(\ref{eq:chialvox})-(\ref{eq:chialvoy}) exhibits chaotic spikes.

\begin{figure}[h]
    \centering
    \includegraphics[width=0.8\linewidth]{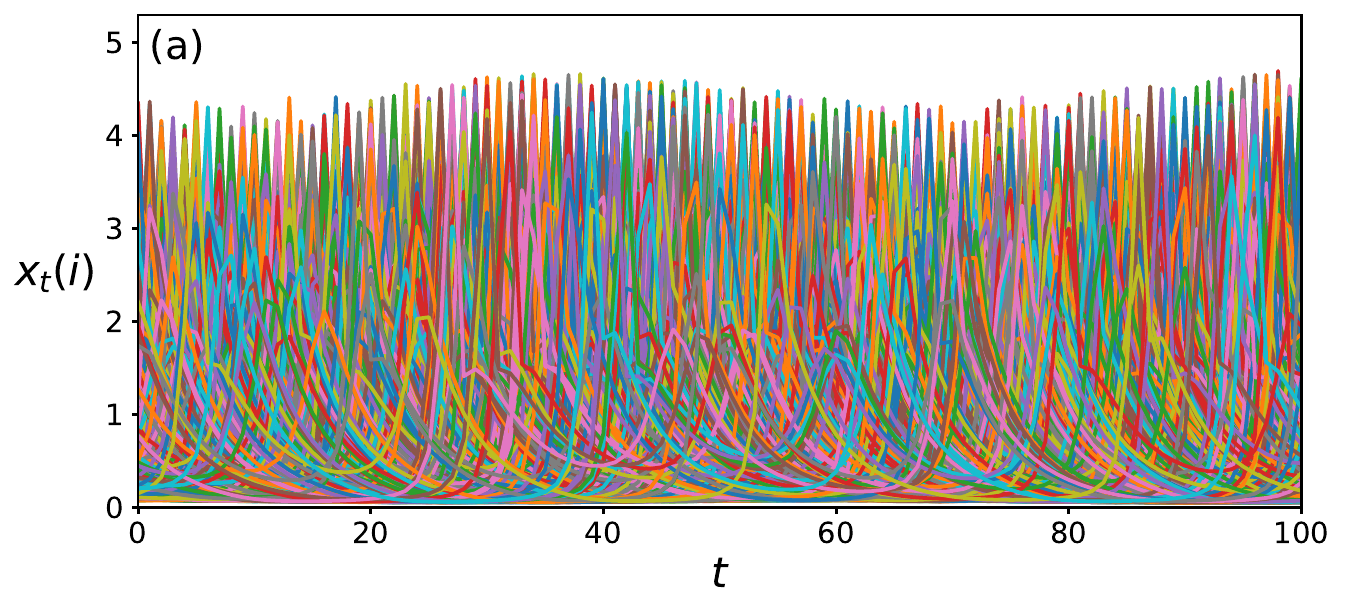}
    \includegraphics[width=0.8\linewidth]{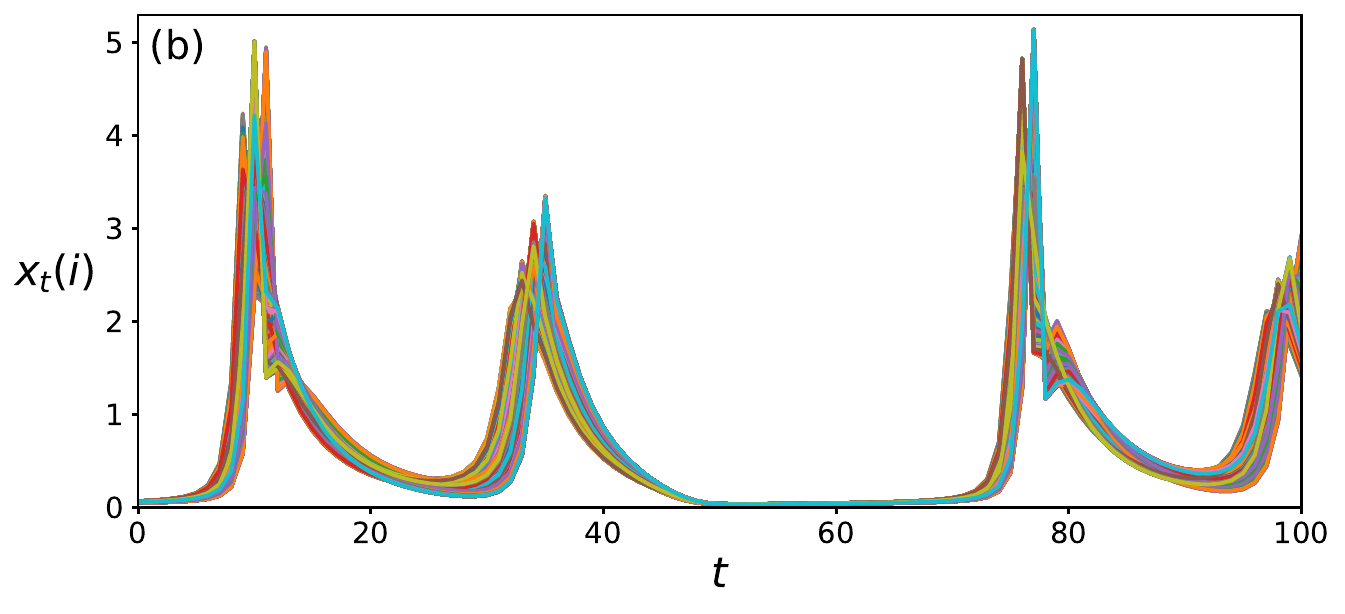}
    \includegraphics[width=0.8\linewidth]{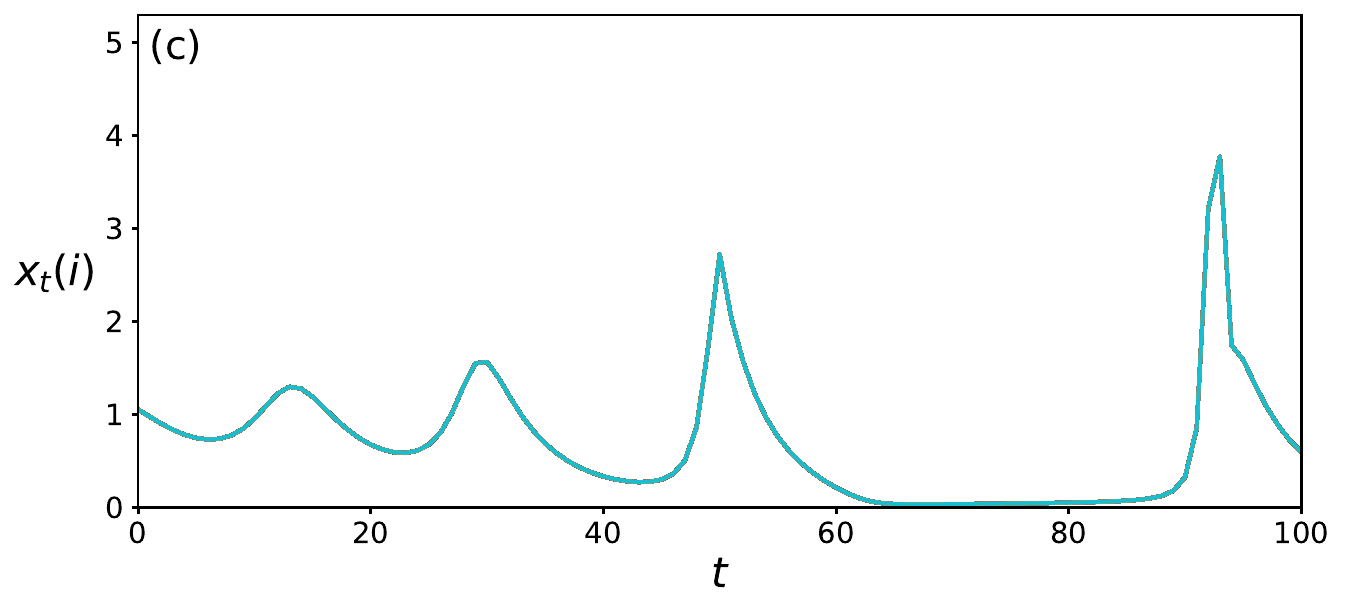}
\caption{Typical time series $x_t(i)$ of  $N=10^3$ maps in the globally coupled neuron system Eqs.~(\ref{eq:first_equation})-(\ref{eq:second_equation}) for different coupling strengths $\epsilon$, after discarding   $5\times10^4$ transients. Fixed map parameters are $a=0.89$, $b=0.18$, and $c=0.28$. Each map $i$ is assigned a heterogeneous parameter $k(i) \in [0.026, 0.03]$.
(a) $\epsilon = 0.028$, desynchronization. (b) $\epsilon = 0.14$, chaotic Griffiths phase.
(c) $\epsilon = 0.7$, chaotic coherence.}
\label{Fig1_timeseries}
\end{figure}

The system Eqs.~(\ref{eq:first_equation})-(\ref{eq:second_equation}) exhibits different behaviors as
the coupling strength $\epsilon$ is varied, as shown in Fig.~\ref{Fig1_timeseries}
where the time evolution of the variables $x_t(i)$ are represented: (a) desynchronization, where the maps are chaotic and desynchronized;
(b) intermittent dynamics, where the maps are chaotic and coherent for a time and subsequently desynchronize in a repetitive process that we shall identify as a chaotic Griffiths phase; and (c) quasi-coherence or coherence, where the states of all the elements evolve closely, forming a single chaotic cluster.
 Note that coherence of the activation variables $x$ also drives
the recovery variables $y$ toward coherence.

\section{Characterizing the chaotic Griffiths phase}
In the chaotic Griffiths phase, all maps 
form a large cluster from time to time,
and then they become desynchronized.
To identify instantaneous clusters, we use a single-linkage procedure. At each time
$t$, the activation states are sorted according to
\begin{equation}
x_t(1)\leq x_t(2)\leq\cdots\leq x_t(N).
\end{equation}
Two consecutive states are linked when
\begin{equation}
x_t(j+1)-x_t(j)\leq\delta_N.
\label{eq:cluster_gap}
\end{equation}
A cluster is a maximal consecutive sequence $j$ satisfying Eq.~(\ref{eq:cluster_gap}), and a new cluster begins whenever the difference between consecutive states exceeds $\delta_N$. 

We denote by $M_t$ the number of nontrivial clusters at time $t$, namely, clusters containing at
least two maps, and by $s_t(\chi)$ the number of maps belonging to
the $\chi$th nontrivial cluster at time $t$.
Then, the size of the largest cluster at time $t$ is $S_{\max}=\max s_t(\chi)$, $\chi=1,2,\ldots,M_t$. 
We calculate the fraction of elements that belong to nontrivial clusters at time $t$ as 
\begin{equation}
p_t=\frac{1}{N}\sum_{\chi=1}^{M_t}s_t(\chi).
\end{equation}
Then, the condition $p_t=1$ indicates that every map belongs
to a nontrivial cluster. A single coherent cluster exists when
$p_t=1$ and $S_{\max}=N$, whereas $p_t=1$ and $S_{\max}<N$ characterize a
multicluster state.
 The incoherent regime is associated with $p_t\simeq 0$, while the chaotic Griffiths phase
displays fluctuating temporal values of $p_t$ in the interval $[0, 1]$.

The clustering threshold $\delta_N$ represents an observational resolution that depends on the system size. If $\delta_N$ is too large, even a desynchronized population appears clustered. If it is too small, even a coherent population is fragmented. As $N$ increases, the typical spacing between neighboring states decreases. 
We calibrate $\delta_N$ by using representative desynchronized (D) and coherent (S) trajectories. We retain the values of $\delta_N$ for which $p_t<0.10$ during at least $95\%$ of the desynchronized trajectory and $p_t>0.90$ during at least $95\%$ of the coherent trajectory.
The thresholds satisfying both conditions define an admissible interval. In this paper,
we set $\delta_N$ to the upper bound of this interval. This selection is systematically applied for every system size $N$. Trajectories in the chaotic Griffiths phase are not used in the calibration.

\begin{figure}[h]
    \centering
    \includegraphics[width=.4\textwidth]{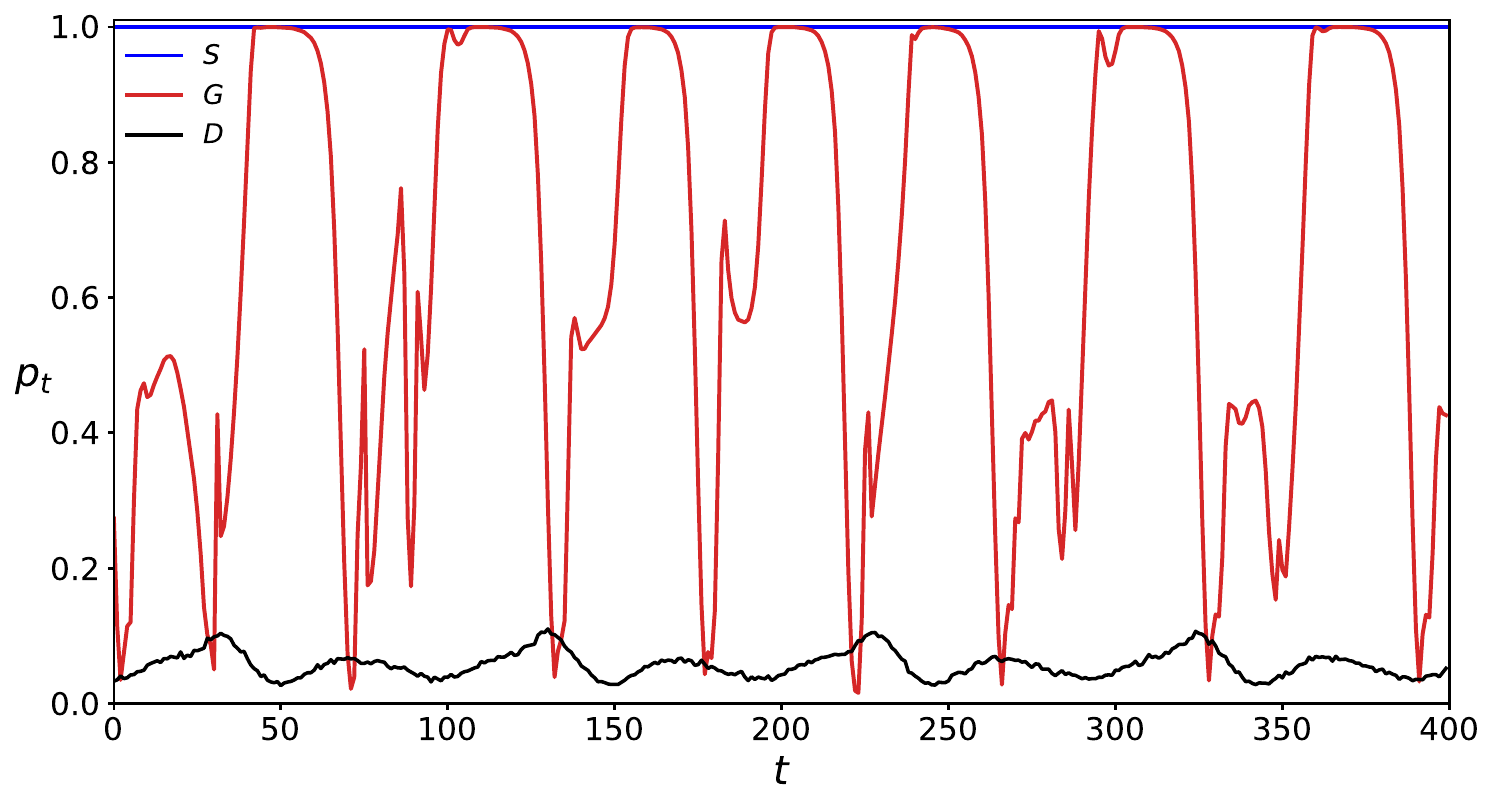}
\caption{Fraction of clustered neurons  $p_t$  as a function of time using a threshold $\delta_{16000}=1.41\times10^{-6}$ for the system
Eqs.~(\ref{eq:first_equation})-(\ref{eq:second_equation}),  for different values of $\epsilon$. A total of $5.0 \times 10^4$ transient iterations have been discarded. Fixed parameters are $a=0.89$, $b=0.18$, $c=0.28$, size $N=16000$, and $k(i) \in [0.026,\,0.03]$.
Black line: $\epsilon=0.028$, D (desynchronization). Red line: $\epsilon=0.14$, G (chaotic Griffiths phase). Blue line: $\epsilon=0.8$, S (coherence).}
    \label{Fig2_Pt}
\end{figure}

Figure~\ref{Fig2_Pt} shows the time evolution of the fraction of clustered neurons, $p_t$, in the system Eqs.~(\ref{eq:first_equation})-(\ref{eq:second_equation}) for
the different behaviors seen in Fig.~\ref{Fig1_timeseries}, denoted by S (coherence), G (chaotic Griffiths phase), and D (desynchronization).
The cluster dynamics of the chaotic Griffiths phase (G) reveals large fluctuations in the quantity $p_t$ over time, alternating
between irregular intervals of coherence and desynchronization. 
The intermittent
coherence-desynchronization process is similar to the phenomenon of chaotic itinerancy found in globally coupled maps \cite{Kaneko2003}, but here this behavior exists on an extended parameter region, rather than only at a critical parameter value.

To quantify the great variability exhibited by $p_t$ in the chaotic Griffiths phase, we
define the dispersion of the  
clustered fraction $\sigma(p)$,
calculated as the asymptotic standard deviation (after discarding a number of transients) of the instantaneous fraction of clustered neurons,
\begin{align}
\sigma (p) = \left[\frac{1}{T} \sum_{t=1}^T \left(p_t - \bar{p}\right)^2 \right]^{1/2} ,
\label{eq:your_label}
\end{align}
where $\bar{p}$ is the asymptotic time-average of $p_t$ defined as
\begin{equation}
    \bar{p} = \frac{1}{T} \sum_{t=1}^T p_t,
\label{average_p}
\end{equation}
for large $T$, calculated after discarding a number of transients. 
Therefore, the chaotic Griffiths phase should be characterized by large values of
$\sigma(p)$ compared with  those in both the coherent and incoherent regimes. In numerical simulations, we consider a threshold
$\sigma(p) > \sigma(p)_{\min}$ to identify the chaotic Griffiths phase.
Then, the desynchronized phase (D) corresponds to $\sigma(p) < \sigma(p)_{\min}$ and $\bar{p} \to 0$,  and the coherent phase (S) is described by $\sigma(p)=0$ and $\bar{p}=1$.
The choice of $\sigma(p)_{\min}$ should accommodate the large variability in the fraction of clustered neurons associated with the chaotic Griffiths phase. 
For the system Eqs.~(\ref{eq:first_equation})-(\ref{eq:second_equation}) we set $\sigma(p)_{\min}=0.12$.

\begin{figure}[h]
\centering \includegraphics[width=.4\textwidth]{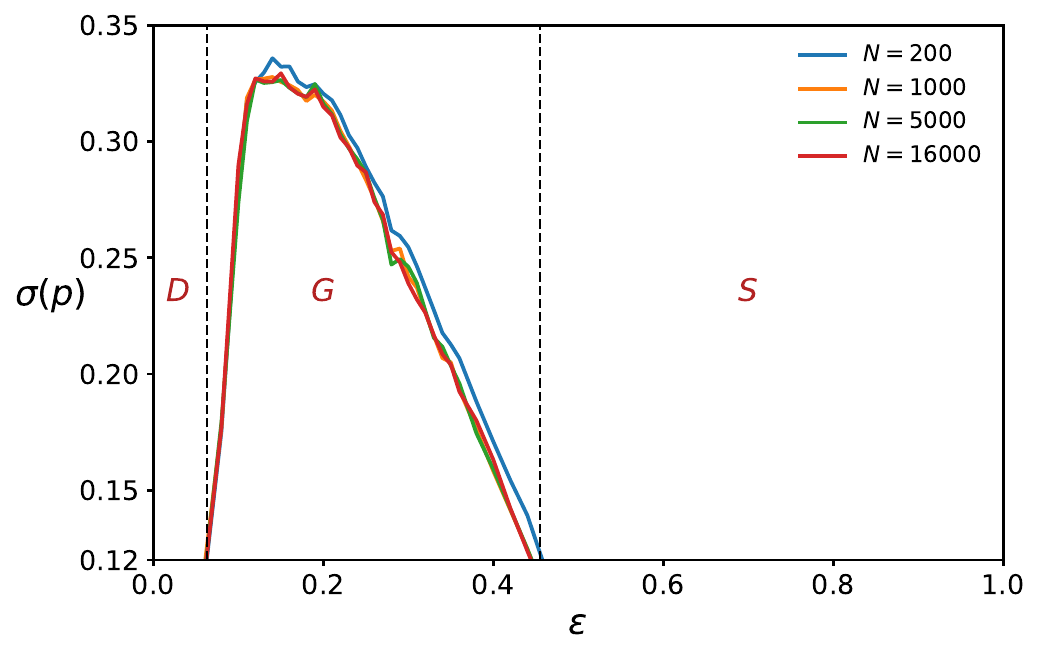} \caption{Dispersion of the clustered fraction $\sigma(p)$ vs coupling $\epsilon$ for the system Eqs.~(\ref{eq:first_equation})--(\ref{eq:second_equation}) and different system sizes $N=200$, $1000$, $5000$, and $16000$. 
Only values $\sigma(p)>0.12$ are plotted. 
For each $N$, clusters are identified using the appropriate threshold $\delta_N$, defined in the text. Fixed parameters are $a=0.89$, $b=0.18$, $c=0.28$, and $k(i)\in[0.026,\,0.03]$. For each value of $\epsilon$, $\sigma(p)$ is averaged over $20$ independent realizations with different random initial conditions and quenched heterogeneity, after discarding $2\times10^4$ transient iterations. Labels indicate D: desynchronization, G: chaotic Griffiths phase, and S: coherence.}
    \label{Fig3_DFC}
\end{figure}

Figure~\ref{Fig3_DFC} shows the  dispersion of the clustered fraction $\sigma(p)$ as a function of
the coupling strength $\epsilon$ for the system Eqs.~(\ref{eq:first_equation})-(\ref{eq:second_equation})  with different sizes. 
The quantity $\sigma(p)$ is greater than the threshold $\sigma(p)_{\min}=0.12$
over a range of values of the coupling strength $\epsilon$, identified as the chaotic Griffiths phase (G). This phase appears as intermediate between the desynchronization phase (D) and the coherent regime (S).

To investigate the relation between the emergence of the chaotic Griffiths phase and local heterogeneity, we assign the map parameters $k(i)$ as
\begin{equation}
k(i) = \frac{(k_1 + k_2)}{2} + A \, \left(\xi(i)-\frac{1}{2} \right) (k_2 - k_1),
   \label{Ap}
\end{equation}
where the factor $A$ varies between $0$ and $1$, $\xi(i)$ is a random number uniformly distributed in $[0, 1]$, and $k_1=0.026$ and $k_2=0.03$  define the interval of the parameter $k$. The coefficient $A$ controls the amount of local heterogeneity; $A=0$ corresponds to a completely homogeneous system ($k(i)=(k_1 + k_2)/2$, $\forall i$), while $A=1$ describes the maximum heterogeneity in the local parameter ($k(i) \in [k_1, k_2]$). Intermediate values of $A$ continuously interpolate between these two limits. 

\begin{figure}[h]
    \centering
    \includegraphics[width=.45\textwidth]{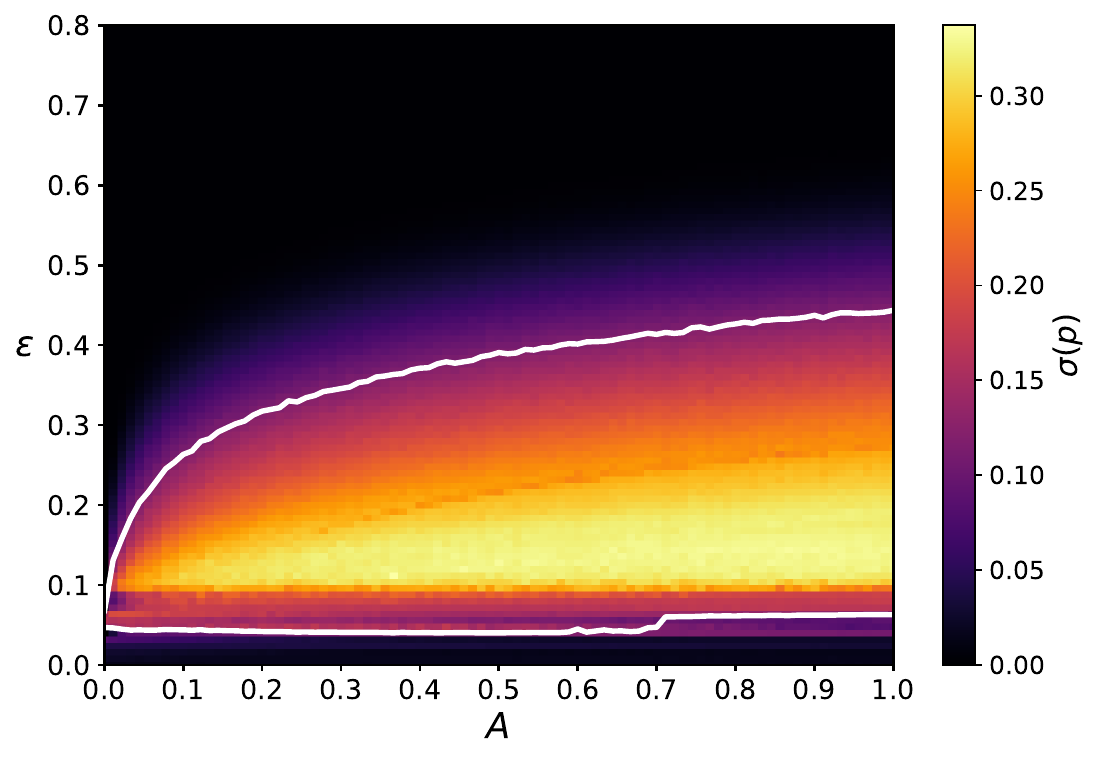}
\caption{Dispersion of the clustered fraction $\sigma(p)$ on the  parameter plane $(A,\epsilon)$, for the system Eqs.~(\ref{eq:first_equation})--(\ref{eq:second_equation}).  The values of $\sigma(p)$ are indicated by the color code bar on the right. The white contour curves enclose the regions of parameters for which $\sigma(p)>\sigma_{\min}=0.12$. Fixed parameters are $N=1000$, $a=0.89$, $b=0.18$, $c=0.28$, and the calibrated threshold $\delta_{1000}=2.2\times10^{-5}$. Values shown are averaged over $20$ independent realizations with different initial conditions.}
\label{Fig_Heterogeneity}
\end{figure}

Figure~(\ref{Fig_Heterogeneity}) shows the dispersion of the clustered fraction $\sigma(p)$ calculated on the space of parameters $(A, \epsilon)$ for the system Eqs.~(\ref{eq:first_equation})--(\ref{eq:second_equation}). The chaotic Griffiths phase, satisfying $\sigma(p)>\sigma_{\min}$, is identified between the white line boundaries in Fig.~(\ref{Fig_Heterogeneity}). 
No  chaotic Griffiths phase appears if the neurons are homogeneous, i. e. $A=0$, while the range of coupling parameter $\epsilon$ where this phase emerges widens with increasing local heterogeneity, reaching the maximum at $A=1$, in agreement with the chaotic Griffiths phase region labeled $G$ in Fig.~\ref{Fig3_DFC}. 

\begin{figure}[h]
    \centering
        \includegraphics[width=.45\textwidth]{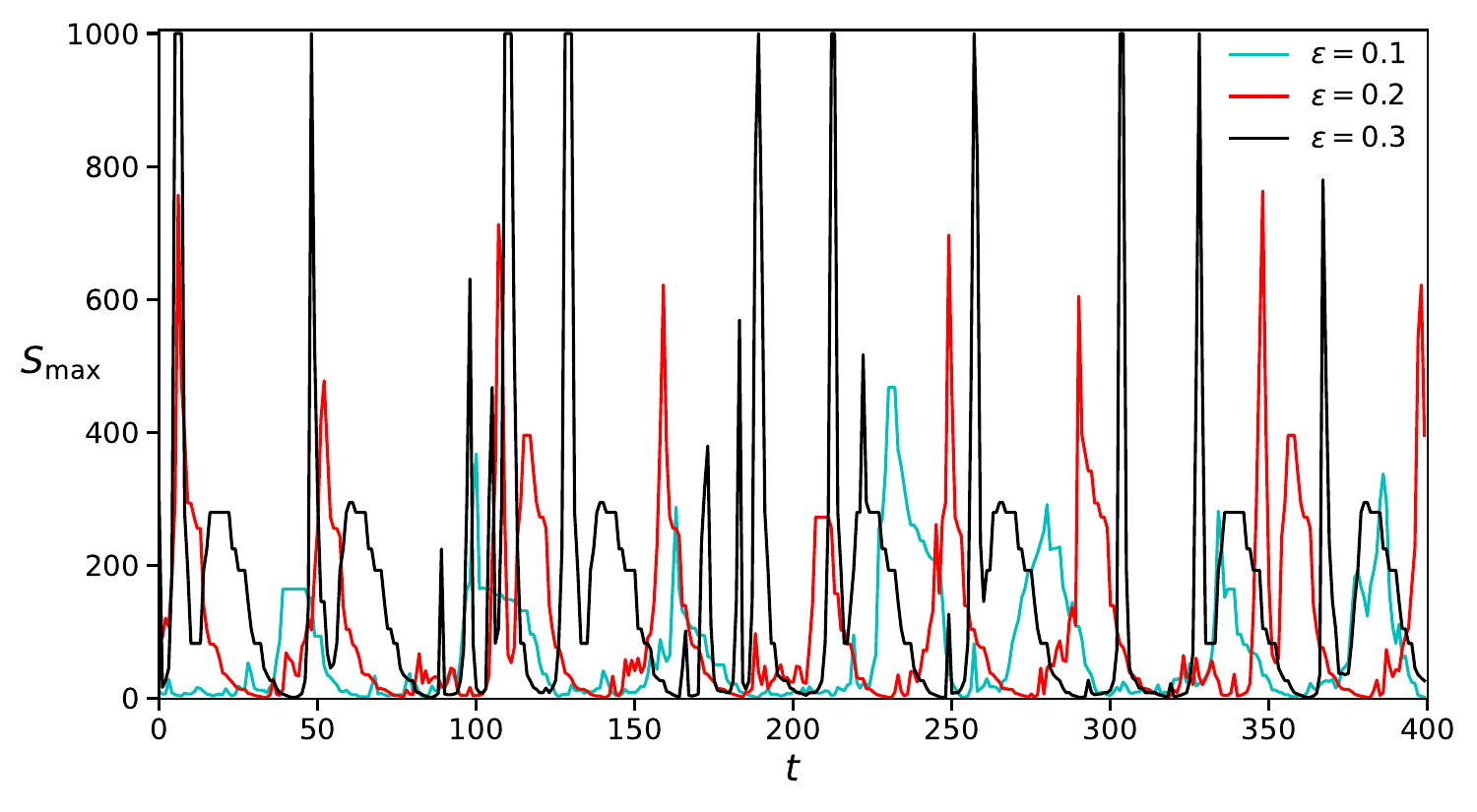}
    \caption{Time evolution of the  maximal cluster size $S_{\max}$ in  the system Eqs.~(\ref{eq:first_equation})-(\ref{eq:second_equation}), for different values of the coupling $\epsilon$, after discarding $5.0\times10^4$ transients. Fixed parameters
$N = 1000$, $a=0.89$, $b=0.18$, $c=0.28$, $k(i) \in [0.026, 0.03]$,
$\delta_{1000}=2.2\times10^{-5}$. Different color lines indicate the coupling strengths within the chaotic Griffiths phase: $\epsilon = 0.1$ (blue), $\epsilon = 0.2$ (red), and $\epsilon = 0.3$ (black).}
    \label{Fig4_Smax}
\end{figure}

The variability in the cluster-formation dynamics characteristic of the
chaotic Griffiths phase is reflected in the behavior of the largest cluster in the system.
Figure~(\ref{Fig4_Smax}) shows the time series of the maximal cluster size $S_{\max}$  in  the system Eqs.~(\ref{eq:first_equation}) and (\ref{eq:second_equation}) for different values of the coupling $\epsilon$ in the region identified as G in Fig.~\ref{Fig3_DFC}.

\begin{figure}[h]
    \centering
\includegraphics[width=.4\textwidth]{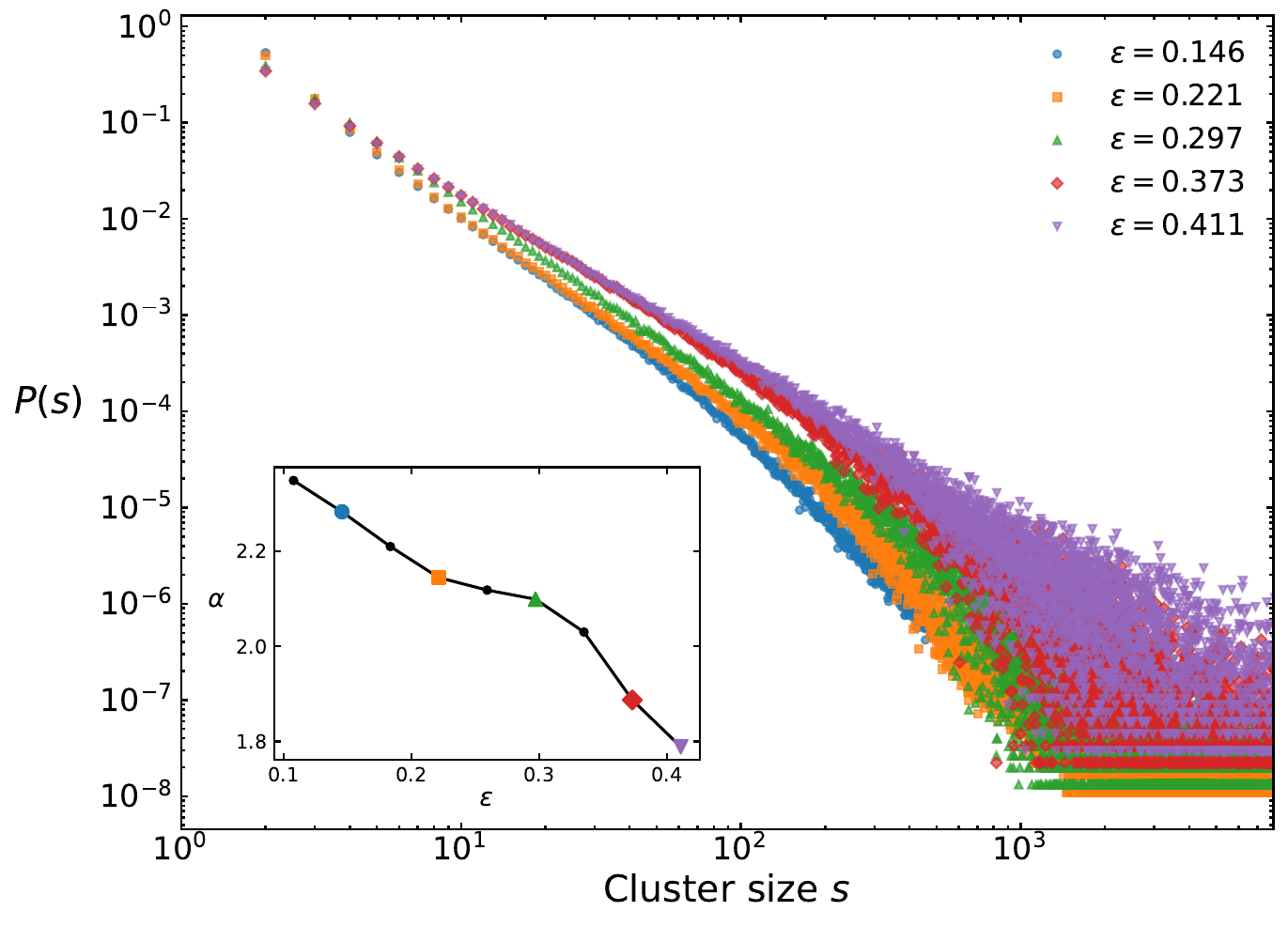}
\caption{Log-log plot of the distribution $P(s)$ as a function of the cluster size $s$ for the system Eqs.~(\ref{eq:first_equation})--(\ref{eq:second_equation}) with different coupling strengths $\epsilon$. Clusters are identified using the upper boundary of the admissible threshold interval, $\delta_{16000}=1.41\times10^{-6}$. Fixed parameters are $N=16000$, $a=0.89$, $b=0.18$, $c=0.28$, and $k(i)\in[0.026,\,0.03]$. Statistics are accumulated over $10^3$ time steps, after discarding $5.0\times10^4$ transient iterations, and averaged over $100$ independent realizations with different initial conditions and quenched parameter distributions. The curves for $\epsilon=0.146$, $0.221$, $0.297$, $0.373$, and $0.411$ are plotted with different colors. Inset: exponent $\alpha$ vs. $\epsilon$.}
    \label{Fig5_Powerlaw}
\end{figure}

A main signature of the Griffiths phase is the emergence of
criticality with power-law statistics over a range of parameters.
Thus, we calculate the probability distribution $P(s)$ of the cluster sizes $s$ obtained from long time series of the system  Eqs.~(\ref{eq:first_equation})-(\ref{eq:second_equation}), for different values of $\epsilon$ in the range denoted by $G$ in Fig.~\ref{Fig3_DFC}. As shown in Fig.~(\ref{Fig5_Powerlaw}), the distribution
$P(s)$ follows a power-law $P(s) \sim s^{-\alpha}$, where the exponent
$\alpha$ changes monotonically with $\epsilon$.
The exponents are calculated using a maximum-likelihood estimator over the cluster size range $s \in [10, 1000]$. 
The scale-free clustering behavior observed in Fig.~(\ref{Fig5_Powerlaw}) is indicative of critical dynamics over an extended range of the coupling $\epsilon$.

\section{Local and topological heterogeneity}

There is substantial evidence that Griffiths phases can result from topological disorder in networks \cite{Munoz2010, juhasz2012rare, Moretti2013, Cota2016}. In particular, the chaotic Griffiths phase was discovered in identical logistic maps coupled on random Erd\"os-R\'enyi networks \cite{Shinoda2016}. Intermittent synchronized and desynchronized dynamics has also been reported in small-world networks of chaotic R\"ossler oscillators \cite{Choudhary2017}. 
However, the emergence of the chaotic Griffiths phase in the globally coupled system  Eqs.~(\ref{eq:first_equation})-(\ref{eq:second_equation}) is a consequence of the heterogeneity in the local neuron maps represented by the distribution of the
parameters $k(i)$. 

Motivated by these findings, we explore the effects of the combined presence of parameter heterogeneity and topological disorder in a network of coupled neurons. We consider a system of Chialvo maps
Eqs.~(\ref{eq:chialvox})-(\ref{eq:chialvoy}) coupled on a  Watts--Strogatz small-world network \cite{Watts1998}, defined as
\begin{eqnarray}
\label{sw1}
x_{t+1}(i) &=&  (1-\epsilon) f\left(x_t(i), y_t(i), k(i) \right) \nonumber \\
&+& \frac{\epsilon}{n(i)} \sum_{j=1}^N  A_{ij} f\left(x_t(j), y_t(j), k(j) \right) ,\\
y_{t+1}(i)&=&g(x_t(i),y_t(i)),
\label{sw2}
\end{eqnarray}
where 
$A_{ij}$ is the adjacency matrix representing the small-world network for a rewiring  probability $r$ and $n(i)$ is the number of neighbors or the degree of node $i$. 
The small-world topology captures features relevant to neuronal networks while introducing structural heterogeneity absent in the globally coupled system.
Equations (\ref{sw1}) and (\ref{sw2}) can be generalized to
include other complex networks or local dynamics.

\begin{figure}[h]
    \centering
    \includegraphics[width=0.4\textwidth]{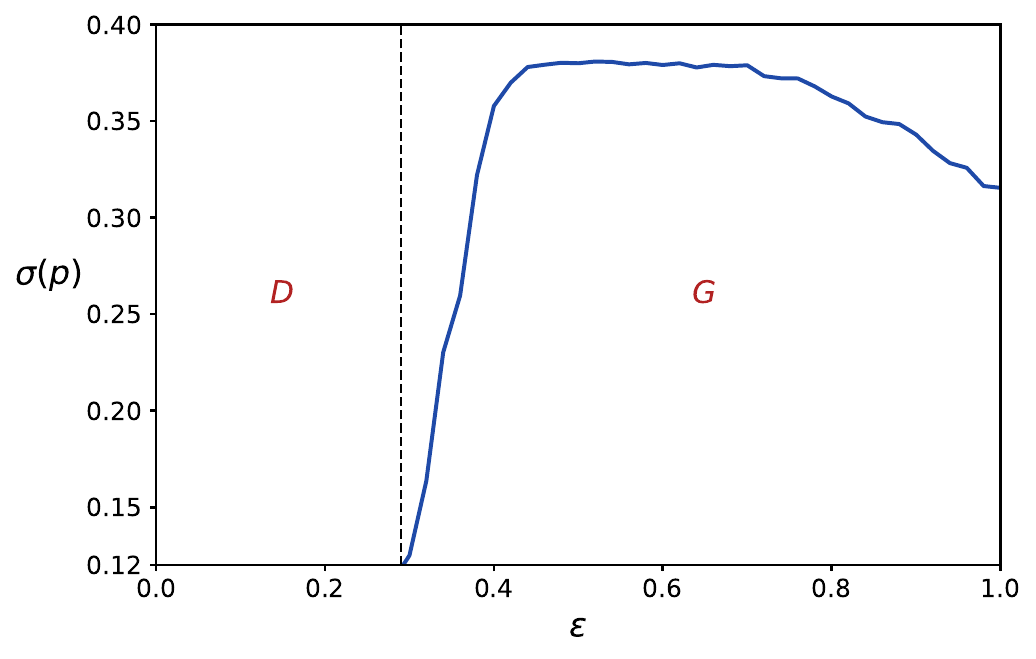}
\caption{Dispersion of the clustered fraction $\sigma(p)$ as a function of the coupling $\epsilon$ for a small-world network of coupled Chialvo maps
Eq.~(\ref{sw1})-(\ref{sw2}). Only values $\sigma(p)$ greater than the threshold $0.12$ are plotted.
Simulations are performed on a  Watts--Strogatz small-world network of size $N=200$ with  $r=0.04$ and  $\bar{n}=40$.
Fixed parameters: $a=0.89$, $b=0.18$, $c=0.28$,  $k(i) \in [0.026,\,0.03]$,
$\delta=10^{-4}$.
For each value of $\epsilon$, the value $\sigma(p)$ shown is the average over $30$ realizations of random initial conditions, after discarding  $5\times10^4$ transient iterations.
Labels indicate D: desynchronization and G: chaotic Griffiths phase.}
    \label{Fig6_DFC}
\end{figure}

Figure~\ref{Fig6_DFC} shows the dispersion of the clustered fraction 
$\sigma(p)$ as a function of the coupling parameter $\epsilon$ for the system
Eqs.~(\ref{sw1}) and (\ref{sw2}) defined on a Watts--Strogatz small-world network with a rewiring  probability $r=0.04$.
The effect of both parameter heterogeneity and structural heterogeneity is  manifested by the extension of the chaotic Griffiths phase  over a broader range of the coupling strength $\epsilon$, effectively
inhibiting the system from reaching a coherent regime.

\begin{figure}[h]
    \centering
\includegraphics[width=0.4\textwidth]{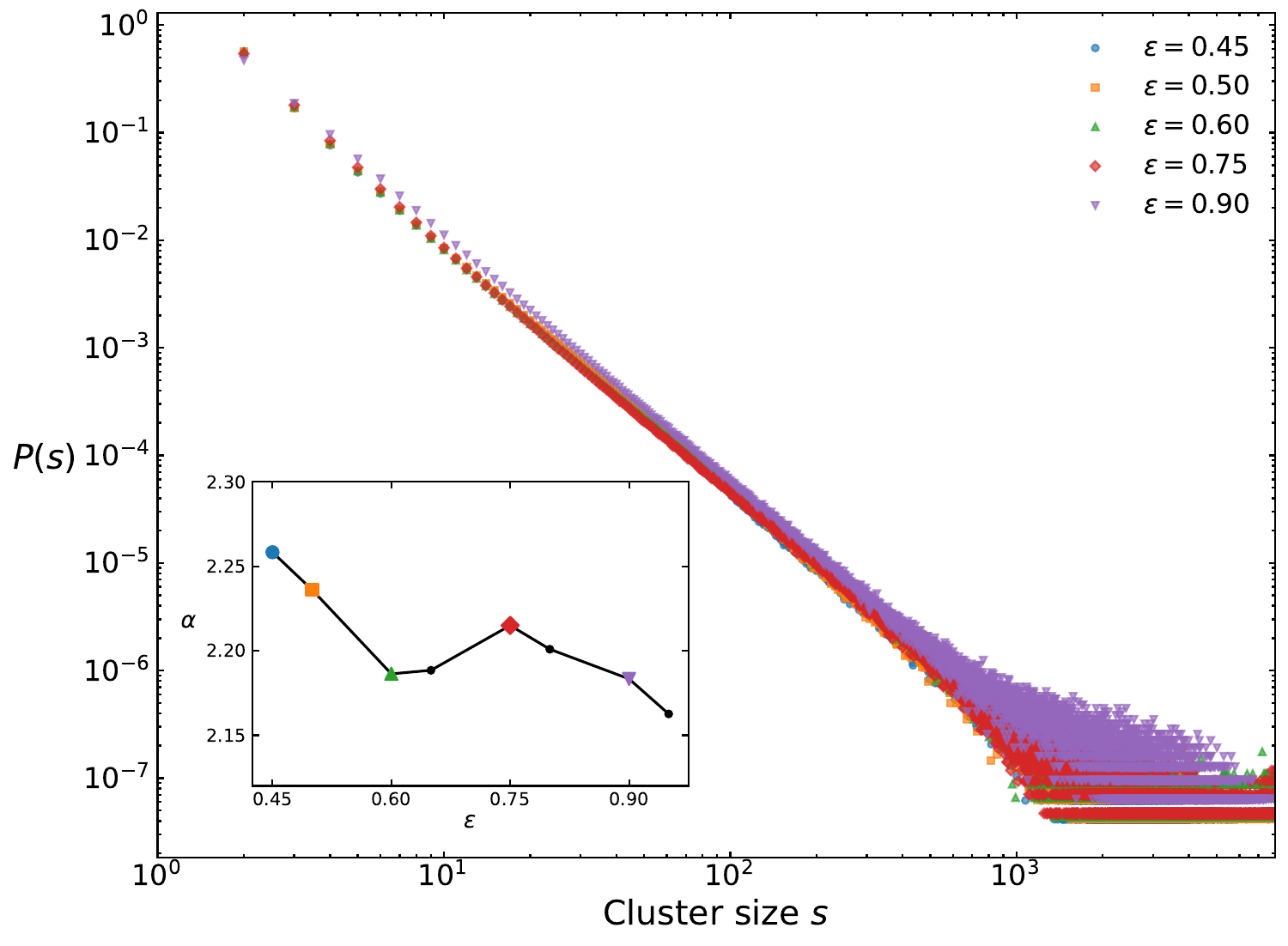}
\caption{Log-log plot of the distribution $P(s)$ as a function of the cluster size $s$ for Eqs.~(\ref{sw1})--(\ref{sw2}) on a Watts--Strogatz small-world network of size $N=10000$, $r=0.04$, and  $\bar{n}=40$. Fixed parameters are $a=0.89$, $b=0.18$, $c=0.28$, $k(i)\in[0.026,\,0.03]$, and a  threshold $\delta_{10000}=2.239\times10^{-6}$. Statistics are averaged over $100$ initial conditions and $100$ independent network realizations, using $10^3$ time steps, after discarding $5.0\times10^4$ transients. The curves for $\epsilon=0.45$, $0.50$, $0.60$, $0.75$, and $0.90$ are plotted with different colors. Inset: exponent $\alpha$ vs. $\epsilon$.}
\label{Fig7_Powerlaw}
\end{figure}

Figure~\ref{Fig7_Powerlaw} provides evidence of criticality in the extended chaotic Griffiths phase region (G) of Fig.~\ref{Fig6_DFC}. Power-law scaling $P(s) \sim s^{-\alpha}$ in the distribution of cluster sizes  is observed over the parameter range $\epsilon \in [0.4, 1]$. The exponent $\alpha$ varies with the coupling strength $\epsilon$.

\begin{figure}[h]
    \centering

    \includegraphics[width=0.3\textwidth]{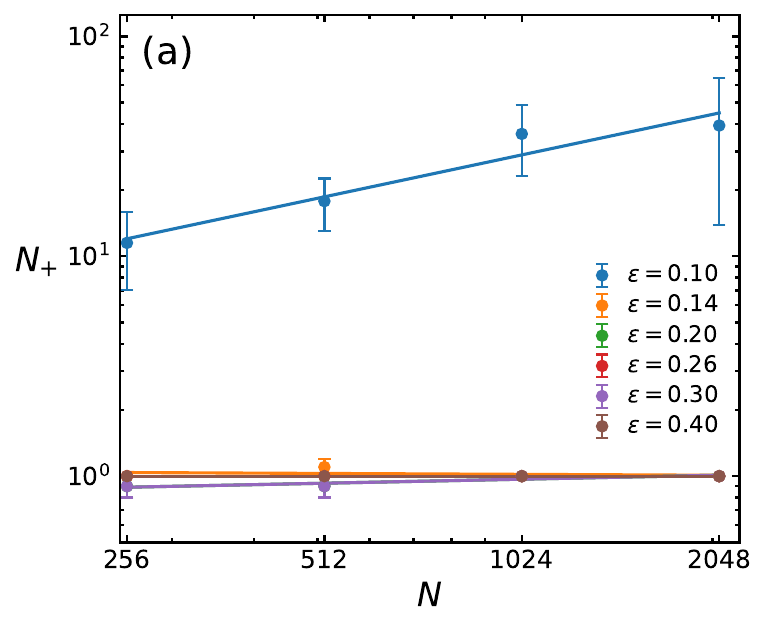}
    \includegraphics[width=0.3\textwidth]{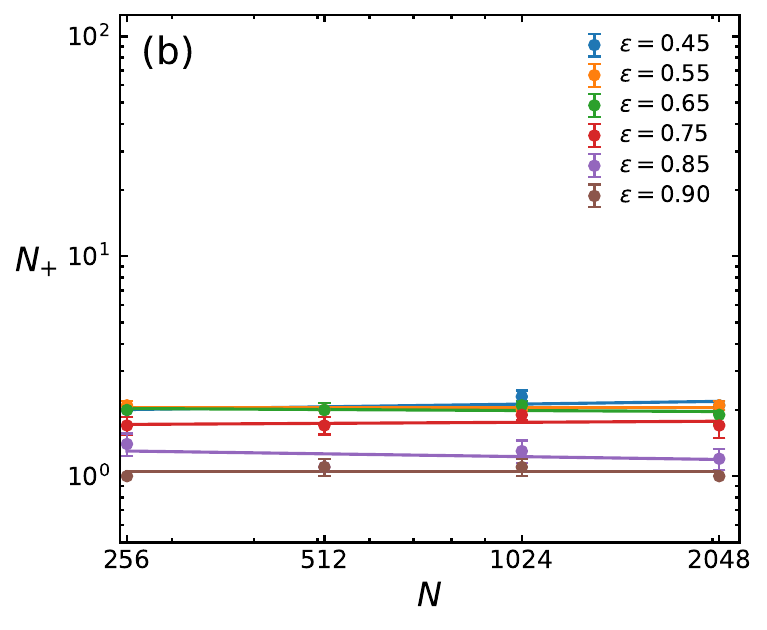}

    \caption{Mean number of positive Lyapunov exponents $N_{+}$ versus system size $N$ for various coupling strengths $\epsilon$ in the chaotic Griffiths phase. For each $(\epsilon,N)$, $N_{+}$ is averaged over $10$ different initial conditions, with error bars indicating the standard error of the mean across realizations. Lyapunov exponents are computed after discarding $5\times10^{4}$ transient iterations of the state dynamics, and are estimated over $2\times10^{3}$ iterations. (a) Globally coupled heterogeneous map network. (b) Small-world heterogeneous map network.}
    \label{Fig9_exponents}
\end{figure}

For spatiotemporal chaos in regular coupled map lattices and in globally coupled map lattices, the number of positive Lyapunov exponents is an extensive quantity that increases
linearly with the system size \cite{Kaneko1986,Takeuchi2011}. 
In contrast, Shinoda and Kaneko \cite{Shinoda2016} found an anomalous subextensive scaling of this quantity in the chaotic Griffiths phase of identical maps coupled on a disordered network, and identified this behavior as a characteristic signature of the phase.
Figure~\ref{Fig9_exponents} shows the mean number of positive Lyapunov exponents, $N_+$, as a function of the size $N$ in the chaotic Griffiths phase for both systems studied here. 
In both cases, we have verified anomalous scaling as $N_+ \sim N^\beta$, with $\beta \approx 0$ for most coupling values. 
Figure~\ref{Fig9_exponents}(a) reveals that subextensive scaling occurs when disorder originates solely from quenched heterogeneity of the local neuronal parameters in the globally coupled network. Thus, topological disorder is not required for the anomalous scaling of the unstable degrees of freedom associated with the chaotic Griffiths phase.
Anomalous scaling is also observed when local parameter heterogeneity and topological disorder are simultaneously present (Fig.~\ref{Fig9_exponents}(b)). 
The resulting anomalous scaling exponents $\beta$ for most coupling parameters are close to zero 
and smaller than those in the previously reported case \cite{Shinoda2016}. 
This difference indicates that, while anomalous Lyapunov scaling appears to be a robust signature of the chaotic Griffiths phase, its quantitative scaling properties depend on the nature of the disorder.
Together with intermittent clustering and power-law cluster size distributions over a finite parameter range, these findings sustain an extension of the chaotic Griffiths-phase scenario to heterogeneous neuronal map networks.

\section{Conclusions}
We have investigated the emergence of a chaotic Griffiths phase in networks of coupled map-based neurons, focusing on the role of quenched heterogeneity in the local dynamics. In contrast with previous studies in which Griffiths phases have been primarily associated with structural disorder, our results show that topological heterogeneity is not necessary for the emergence of a chaotic Griffiths phase. In a globally coupled network of Chialvo maps, quenched disorder in a local neuronal parameter is sufficient to produce an extended intermediate regime between desynchronized and coherent chaotic dynamics. Moreover, the chaotic Griffiths phase is absent for identical neurons and progressively broadens as the degree of local parameter heterogeneity increases, establishing a direct connection between intrinsic dynamical diversity and the emergence of extended critical behavior. 

The chaotic Griffiths phase is characterized by the intermittent formation and destruction of neuronal clusters and large temporal fluctuations in the fraction of clustered neurons. The cluster-size distributions exhibit power-law behavior over a finite interval of coupling strengths, with exponents that vary with the coupling parameter. These results provide evidence that the observed intermittent regime represents extended critical behavior rather than a conventional transition occurring at an isolated critical point. 

We have also explored the combined effects of intrinsic and structural heterogeneity by considering Chialvo maps with heterogeneous local parameters coupled through a small-world network. The introduction of topological disorder substantially extends the range of coupling strengths over which the chaotic Griffiths phase occurs and prevents the system from reaching the coherent regime over the parameter interval investigated. Power-law cluster-size distributions persist throughout this enlarged region. Thus, local dynamical heterogeneity and structural disorder provide distinct but complementary mechanisms for sustaining extended critical behavior.
 
For both the globally coupled heterogeneous network and the system containing parameter and topological disorder, the number of positive Lyapunov exponents exhibits anomalous scaling,
previously identified as a characteristic signature of the chaotic Griffiths phase in topologically disordered networks of identical maps. Here, this scaling occurs even when disorder originates solely from heterogeneous local parameters and persists when both sources of disorder are present. Although anomalous Lyapunov scaling is a robust signature of the chaotic Griffiths phase, its quantitative properties depend on the nature of the disorder.
 
These results extend the chaotic Griffiths-phase scenario from networks with structural disorder to systems with intrinsic dynamical disorder. They show that heterogeneity of the constituent units can provide an independent route to extended criticality and that its combination with heterogeneous connectivity can further enhance this regime. Since neuronal systems naturally display both cell variability and heterogeneous connectivity, our results suggest that the coexistence of multiple forms of quenched disorder may contribute to the robustness of critical-like collective dynamics in neuronal networks.

\section*{Acknowledgments}
M. G. C. acknowledges support from ViceRectorado de Investigaci\'on and Innovaci\'on, Universidad Yachay Tech, Ecuador.

\end{document}